\documentclass[twocolumn]{aastex62}

\usepackage{graphicx}	
\usepackage{amsmath}	
\usepackage{amssymb}	
\usepackage{subfigure}
\usepackage{booktabs}
\usepackage{longtable}
\usepackage{array}
\graphicspath{{./}{figures/}}

\received{2019 April 30}
\accepted{2019 May 29}
\submitjournal{PASP}

\shorttitle{Observations of KIC 8462852}
\shortauthors{Hitchcock et al.}

\begin{document}

\correspondingauthor{James Hitchcock}
\email{james.hitchcock.17@ucl.ac.uk}
\author{James Hitchcock}

\thanks{
\begin{minipage}[t]{0.45\textwidth}
The author is currently affiliated with the University of St. Andrews.
\end{minipage}}
\affil{Department of Physics and Astronomy, University College London, Gower Street, London, WC1E 6BT, United Kingdom}

\author{Stephen J. Fossey}
\affiliation{Department of Physics and Astronomy, University College London, Gower Street, London, WC1E 6BT, United Kingdom}
\affil{University College London Observatory, 553 Watford Way, London, NW7 2QS, United Kingdom}

\author{Giorgio Savini}
\affiliation{Department of Physics and Astronomy, University College London, Gower Street, London, WC1E 6BT, United Kingdom}
\affil{University College London Observatory, 553 Watford Way, London, NW7 2QS, United Kingdom}

\title{Non-grey month-long brightening of KIC 8462852 in the immediate aftermath of a deep dip}



\begin{abstract}
We present an analysis of the results of long-term multi-band photometric monitoring of the enigmatic star, KIC 8462852. Observations in the $B$, $g'$, $V$, $r'$ and $I_C$ passbands have been acquired at University College London Observatory (UCLO) between May 2017 and September 2018. We interrogate the wavelength dependence of the $\sim$month-long dimming and brightening exhibited by the target star over an 85-day interval, immediately following a days-long $\sim5$ per cent drop in brightness on Julian Date (JD) 2458203. Between JD 2458215--300 we measure brightness variations which correspond to relative extinctions of $A_B/A_V = 1.39\pm0.27$, $A_{g'}/A_V = 1.16\pm0.11$, $A_{r'}/A_V = 0.80\pm0.25$ and $A_{Ic}/A_V = 0.49\pm0.19$, from which we infer an Angstrom absorption coefficient of $1.33\pm0.43$ ($R_V \simeq 3.2^{+2.6}_{-1.0}$). As with the days-long `dips', the wavelength dependence of the longer-term brightness variations must also be associated with extinction arising from a dust distribution containing a substantial fraction of sub-micron-sized grains. This implies some common mechanism is responsible for the star's variability over both these short and longer timescales.
\end{abstract}




\section{Introduction}
\label{sec: introduction}

Since its discovery by citizen scientists in \textit{Kepler} data, KIC 8462852 has gained a reputation as one of the most unusual stars in the
Galaxy.
In their discovery paper, \citet{boyajian2016planet} describe this F-type main sequence star's unprecedented behavior, in which it was shown to undergo `dips' in its flux over day-long timescales of up to 22 per cent. Through follow-up ground-based multi-band photometric monitoring, it is now known that the dips have a wavelength dependence associated with extinction from astrophysical dust with a substantial fraction of submicron sized particles ($ < 0.3 \mu m$ in diameter) \citep[e.g.][]{boyajian2018first, deeg2018non}.

A study of archival photographic plates by \citet{schaefer2016kic} found that KIC 8462852 has faded in the $B$-band at an average rate of $0.164 \pm 0.013$ magnitudes per century between 1890 and 1989. Variations in flux over year-long timescales have been measured by both \citet{meng2017extinction} and \citet{davenport2018galex}, who measured a fairly neutral extinction of $R_v \geqslant 5$ and $R_V = 5.0 \pm 0.9$ respectively. More recently, \citet{schaefer2018kic} have shown that colour-dependent variations in flux of a few per cent occur on month-long timescales. From a total of 1866 \textit{BVRI} nightly magnitudes binned over 20-day intervals, an extinction law steeper than $R_V\sim 5$ is inferred, and is found to be consistent with that of canonical ISM extinction ($R_v = 3.1$), again suggestive of sub-micron-sized grains. This is an average value to describe the entire 2.43-year-long \textit{BVRI} light curve, and suggests that the material responsible for both the dips and longer-term dimming have a common origin.

\citet{bodman2018variable} note that no single wavelength dependence describes the series of dipping events of May--September 2017. The authors claim there is a tentative detection of non-grey long-term dimming on which the dips are superimposed. It is suggested that the chromaticism associated with this longer-term variation becomes increasingly neutral as the dip complex progresses, and the extinction is markedly steeper during the first dip (\textit{Elsie}) than the fourth and final dip (\textit{Angkor}) four months later.

In March and April of 2018, ground-based photometric observations of the ongoing Las Cumbres Observatory monitoring campaign revealed KIC 8462852 to undergo two further dips (labelled {\textit{Caral Supe} and \textit{Evangeline}) of $\sim5$ per cent, the deepest seen since the \textit{Kepler} mission \footnote{https://www.wherestheflux.com/blog/page/4}. Since May 2017, we have been carrying out multi-band photometric monitoring of KIC 8462852 using observations from the University College London Observatory (UCLO). In particular, our coverage of $BVg'r'I_c$ photometry in the 85 days following the \textit{Caral Supe} and \textit{Evangeline} dips
allows an investigation into the nature of the material assumed to be trailing the object(s) associated with these dips. The paper is laid out as follows: in Section~\ref{sec: methods} we report our observations and photometric calibrations to transform our photometry to the $BVI_C$ and $g'r'$ standard systems; section~\ref{sec: results} presents our results, in which we demonstrate the chromatic dependence of the extinction over this 85-day interval, and obtain a value for the Angstrom absorption coefficient, $\alpha$, and the ratio of total-to-selective extinction, $R_V$, for the inferred population of dust grains. Section~\ref{sec: conclusions} discusses our conclusions.

\section{Methods}
\label{sec: methods}

\subsection{UCLO observations}
\label{sec:phot@uclo}

Observations of KIC 8462852 at UCLO span from May 2017 to September 2018. Two Celestron C14 (0.35-m) robotic telescopes were used for this observing campaign, and observations were made through a total of eight filters: `Green' (as a proxy for $V$) and Astrodon $R_C I_C$ filters on the C14 West telescope (with SBIG STL6303E CCD), and Astrodon $BVg'r'i'$ filters on the C14 East (with FLI PL9000 CCD), amounting to more than 400 mean magnitudes across all filters.
Typically, ten images were obtained in each filter on each night, with exposure times ranging from 50--75 seconds per image. All images were processed with standard bias, dark, and flat-field reductions each night.

\subsection{Photometry}
\label{sec:phot2}

Our approach to obtaining adequate precision in the long-term photometry is outlined below. 

Photometric measurements of KIC 8462852 and reference field stars were made using the Source Extractor ({\sc SExtractor}) software \citet{bertin1996sextractor}, with automated Kron apertures being preferred; we have found that the use of Kron apertures in our images yields marginally better statistics than fixed-radius circular apertures, probably due to a non-circular aperture adapting to variations in the stellar point-spread-function across the field.

Our observing strategy ensured that for each telescope series, the target and reference stars were located in the same position on the chip each night, within the precision of the `plate-solve-and-guide' robotic time-series. There was an additional field rotation due to a meridian flip of the German equatorial mount, depending on the target hour angle. We checked a sample of transformed target magnitudes (see Section \ref{sec:transform}) obtained either side of a meridian flip, and confirmed they were consistent to within a few millimagnitudes (and within uncertainties), on each telescope, for a range of filters.

We ensured that we had sufficient total integrated flux to obtain the required statistical precision within a single night's observations: typical integrated fluxes of 1--2\,$\times 10^5$ photoelectrons
were obtained in KIC 8462852, per image, yielding $>10^6$ photoelectrons per night in each filter. 

Variations in the flux of KIC~8462852 are of the order of at least 1 per cent for both the short-timescale dips and longer-timescale dimming and brightening. As such, the required precision of the photometry to establish statistically significant variations in the target is at the millimagnitude (mmag) level. The intrinsic photometric uncertainty was typically 4--5 mmag for a single frame, dominated by Poisson noise on the star signal and scintillation noise (given the relatively short exposure times). We averaged the magnitudes derived from multiple (typically, ten) images in a single night, but clipped to exclude $>5\sigma$ outliers --- avoiding the effects of, e.g., bad/hot pixels, cosmic rays --- such that the statistical uncertainty can be reduced to achieve an internal precision of 1--2 mmag per night for bright sources central to the CCD. 

We evaluate the effect of systematic errors which can arise in ground-based photometry (e.g., see \citet{schaefer2018kic}) by examining the long-term stability of the reference-star magnitudes in our own data --- both to select stable, non-variable comparison stars for the photometric calibration; and also as a check on the final residual photometric uncertainties from night to night (\S\ref{sec:n2n}).

\subsection{Transformation to standard passbands and reference star selection}
\label{sec:transform}

Photometric calibration of KIC~8462852 was achieved with reference to bright, unsaturated, nearby standard stars in the field for which standard magnitudes were available in the APASS catalogue \citep{henden2016vizier}. {\citet{lahey2017improving} measure a superior photometric precision in the $V$ and $I_C$ bands for many of the reference stars used in the calibration, and where available, we preferred these measurements over the APASS values.}

{In order to correct for the wavelength-dependent mismatch between our instrument system and the standard passbands, and time-variable effects of atmospheric extinction, transformation of our instrument magnitudes at a given airmass, $X$, $m_{\rm inst}(X)$, into the appropriate standard-system passband, $m_{\rm std}$ (as provided by either APASS or \citet{lahey2017improving}), was achieved through the equation}
\begin{equation}
    m_{\rm inst}(X) - m_{\rm std} = z + \epsilon C_{\rm std}, 
	\label{eq:CC}
\end{equation}

{where $C_{\rm std}$ is a star colour in standard passbands, such as $(B-V)$, and $z$ and $\epsilon$ are the transformation coefficients derived from a weighted fit to standard-star magnitudes and colours (as in the example shown in Fig.~\ref{fig:Calibration}). The standard star colours used in equation~\ref{eq:CC} depend on the bandpass transformation: for $B,V$, and $g'$, APASS $(B-V)$ colours (\citet{henden2016vizier}) were used; for $r'$ and $I_c$, $(r'-I_c)$ colours derived from the APASS and \citet{lahey2017improving} values were used.

{The reference stars used in the calibrations for all filters are the most stable stars from an initial sample of 20 candidate reference stars; we derived a standard magnitude for each reference star, and examined their stability over the entire observing campaign --- for each filter, the stars which had a standard deviation exceeding 10 mmag were iteratively removed, and the calibration was re-performed with the improved sample, until all calibration stars were found to be stable at better than $10$ mmag precision over the entirety of the observing campaign, and across all passbands presented in this work.} The standard deviation of their $\sim$1.3-year-long time series is typically better than 5~mmag; for bright sources, central on the CCD, their standard deviation was as low as 1--2 mmag in all passbands, and the level of the within-night variation measured for these stars can be $<1$ mmag. Finally then, up to eight, and no less than five, reference stars were ultimately used for the transformations, dependent on the filter (see Table ~\ref{table:ref-stars}).

The final uncertainty used to weight each reference star in the fits to equation~\ref{eq:CC}, in each filter, was calculated from the orthogonal sum of its standard deviation in the campaign time series and the standard error on its mean magnitude within each night's data}.

{Substitution of these values in place of the `external' uncertainties provided in the APASS catalogue was also motivated by inspection of the reduced chi-squared values in the fits to equation~\ref{eq:CC}, which were much closer to unity.}

\begin{table}[!htpb]
\begin{center}
\begin{tabular}{ r  c  c  c }
 \hline\hline
Telescope/Filter\textsuperscript{a} & Exp. time & Standard & $N({\rm ref})$\textsuperscript{c} \\ [0.5ex] 
\mbox{} & (seconds) & System \textsuperscript{b} & \\
 \hline
 C14 West / \hfill Green & 60 & $V$ & 7 \\
 $R_C$ & 50 & $r'$ & 8 \\
 $I_C$ & 75 & $I_C$ & 8 \\
 \hline
 C14 East / \hfill $B$ & 75 & $B$ & 6 \\
 $g'$ & 50 & $g'$ & 6 \\
 $V$ & 50 & $V$ & 6 \\
 $r'$ & 50 & $r'$ & 5 \\
 $i'$ & 75 & $I_C$ & 5 \\ [1ex]
\hline
\end{tabular}
\caption{Observational and calibration information for all telescopes/filters used here. $(a)$ The telescope/filters used in this observing campaign; $(b)$ the standard photometric systems to which the measured instrument magnitudes in each filter are transformed (equation~\ref{eq:CC}); $(c)$ the number of reference stars used for the transformation.}
\label{table:ref-stars}
\end{center}
\end{table}

\begin{figure}[!htpb]
	\includegraphics[width=\columnwidth]{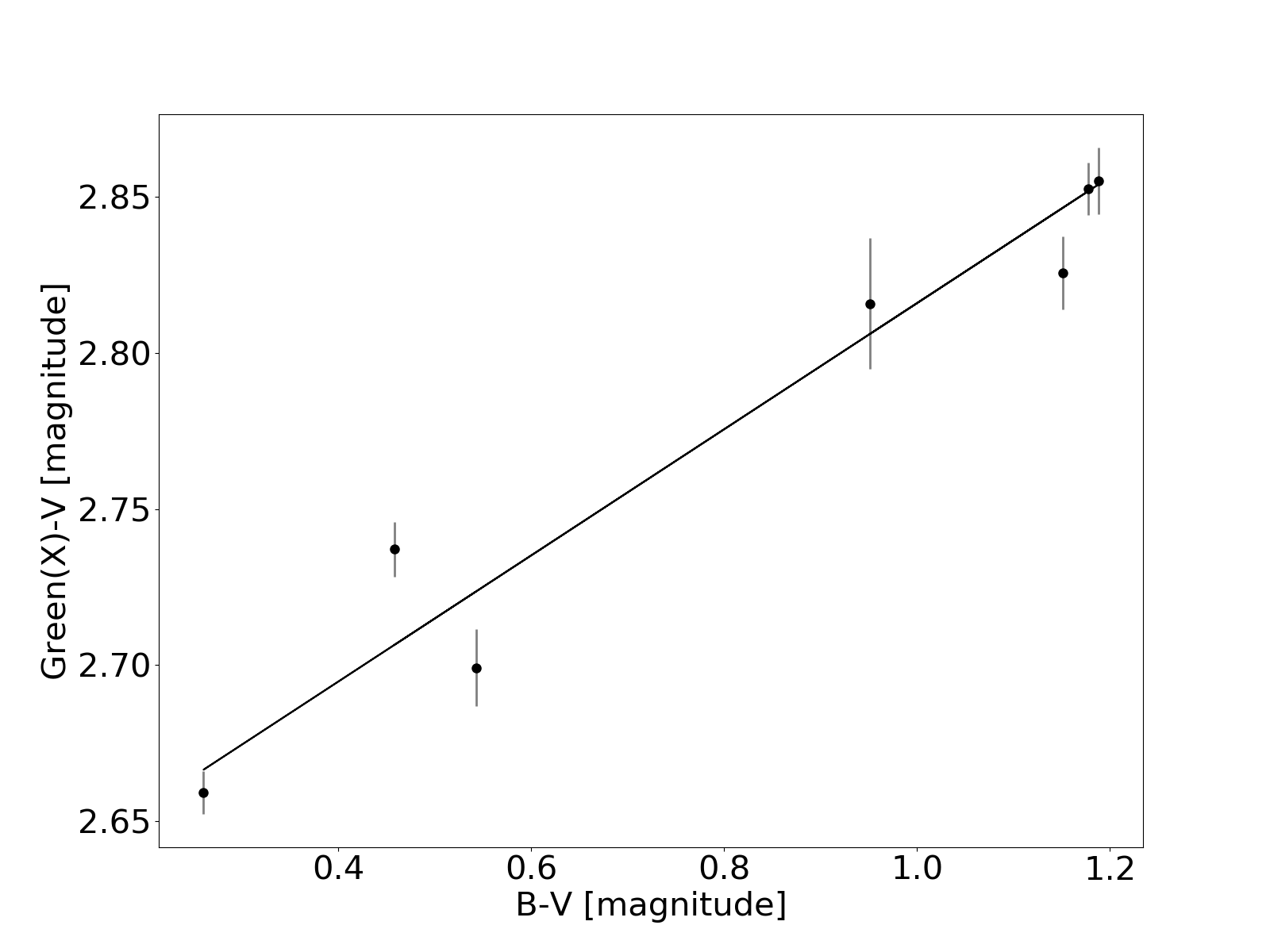}
    \caption{A typical scatter plot from the derivation of the  transformation coefficients (i.e, the intercept, $z$, and slope, $\epsilon$, in equation ~\ref{eq:CC}). This particular plot illustrates the calibration of an instrument magnitude in the C14 West \textit{Green} filter (denoted as a function of airmass, $X$) transformed to the standard $V$ passband.
    }
    \label{fig:Calibration}
\end{figure}

Finally, from the transformation coefficients derived for each image, the instrument magnitudes for KIC~8462852 were transformed to standard magnitudes using equation~\ref{eq:CC}, for each passband indicated in Table~\ref{table:ref-stars}. The standard colours used for KIC~8462852 were $(B-V) = 0.508$ (\citet{henden2016vizier}) and $(r'-I_C) = 0.487$ (based on \citet{henden2016vizier} and \citet{lahey2017improving}) (we always used at least one reference star with a colour at least as blue as that of KIC~8462852 to help establish a reliable colour term in equation~\ref{eq:CC}).

\subsection{Night-to-night uncertainties for KIC 8462852}
\label{sec:n2n}

As noted in \S\ref{sec:phot2}, we can determine intrinsic `within night' uncertainties for the magnitude of KIC~8462852. To account realistically for potential systematic variability from night to night (and hence guard against over-interpretation of brightness variations), we attempt to quantify a systematic uncertainty on each night associated with its transformed standard magnitude. This may be derived from the nightly scatter of the reference-star magnitudes: for any given night, one can calculate the median absolute deviation (MAD) of the reference-star magnitudes, in each filter, relative to their median brightness in the campaign time series, to derive an uncertainty per filter, per night. For night $i$, and reference star $j$,
\begin{equation}
    \sigma_i = 1.4826 \times {\rm median}(|m_{ij} - m_{j,{\rm median}}|).
	\label{eq:MAD}
\end{equation}
Equation~\ref{eq:MAD} hence describes a residual systematic uncertainty, $\sigma_i$, in the standard star magnitudes, for each filter per night, which is combined quadratically with the standard error on the mean of the KIC~8462852 standard magnitudes obtained on the same night. 
These total per-night uncertainties for KIC~8462852 are those shown in Figure \ref{fig:AllMags}, and were used in the weighting of the fits illustrated in Figure \ref{fig:RegPlots}.

\subsection{Combining C14 West and C14 East magnitudes} \label{CombiningC14E+C14W}

{Systematic differences may also occur between observations of the target acquired in different filters on different telescopes, for which we have used different comparison stars. In order concurrently to analyse filter datasets that have been transformed to the same standard photometric bandpass, we renormalise one set of telescope magnitudes to another, separately for $V$, $r'$, and $I_C$ (Table~\ref{table:ref-stars}). This is done by evaluation of the function:}
\begin{eqnarray}
	f = & \sum_i^{N_a} \sum_j^{N_b} \frac{(y_{a,i} - (y_{b,j} + \Delta m))^2}{\sigma_{a,i}^2 + \sigma_{b,j}^2} \qquad  \\ \nonumber
	& \qquad \times \exp\left[-\frac{(t_{a,i}-t_{b,j})^2}{2l^2}\right],
    \label{eq:f}
\end{eqnarray}
(after \citet{osborn2019pds}), where $t_a, t_b$ and $y_a,y_b$ represent times (JD) and magnitudes for series $a$ and $b$, comprising $N_a$ and $N_b$ data points, with magnitude uncertainties $\sigma_a$ and $\sigma_b$. The constant $l$ is fixed to represent the typical shortest-timescale variation of KIC 8462852, and is set here to be 1~day. The magnitude offset $\Delta m$ is optimised for a given filter, by minimising the function $f$. In this way, a single dataset containing both C14 East and renormalised C14 West observations may be used in the analysis. {We chose to renormalise C14 West measurements to the C14 East measurements for all relevant filter datasets, owing to the latter's greater coverage when both telescopes were operational in the campaign (i.e., the observing window which follows the May 2017 dip events, see Figure ~\ref{fig:AllMags}).} For the transformed $V$, $r'$ and $I_c$ measurements obtained with the C14 West, these renormalising offsets are -0.015, +0.032 and +0.006 magnitudes respectively.

A full table of UCLO observations of KIC 8462852 --- but \textit{before} renormalising the C14 West magnitudes --- is publicly available for download in machine-readable form on Figshare\footnote{\label{figfoot}https://figshare.com}. An extract of the first and last three lines of this table is shown in Table \ref{AllMagsTab}.

\section{Results and Discussion}
\label{sec: results}

The UCLO light curve from May 2017 to September 2018 for all passbands used in this work is shown in Fig.~\ref{fig:AllMags}. As in the \citet{schaefer2018kic} light curve --- which spans the the JD window 2457300--8200 --- the amplitude of variation in $B$ can be seen to exceed that in all other bands. Unfortunately, a gap in UCLO coverage means we missed the \textit{Caral-Supe} dip and the minimum of the \textit{Evangeline} dip. However, we do capture the egress of the \textit{Evangeline} dip; over the 85-day window immediately following \textit{Evangeline}, KIC 8462852 undergoes a net brightening, interspersed with shorter-timescale variation, shown as the shaded region in Fig.~\ref{fig:AllMags}. In the following, we interrogate the wavelength dependence associated with this 85-day interval. 

\begin{table*}[!htpb]
    \centering
    \begin{tabular}{>{\centering\arraybackslash}l  >{\centering\arraybackslash}p{0.15\textwidth} >{\centering\arraybackslash}p{0.15\textwidth} >{\centering\arraybackslash}p{0.15\textwidth} >{\centering\arraybackslash}p{0.15\textwidth} }
    
     \hline\hline
    Julian date & Passband & Telescope & Magnitude & Magnitude error \\
     \hline
    
    2458039.49 & $B$ & C14E & 12.4057 & 0.0045 \\
    2458106.29 & $B$ & C14E & 12.4297 & 0.0061 \\
    2458137.76 & $B$ & C14E & 12.4171 & 0.0035 \\
    ... & ... & ... & ... & ... \\
    2458352.44 & $I_c$ & C14W & 11.2022 & 0.0016 \\
    2458354.57 & $I_c$ & C14W & 11.1966 & 0.0019 \\
    2458365.50 & $I_c$ & C14W & 11.2069 & 0.0015 \\
    
    \hline
    \end{tabular}
    \caption{An extract of 1.3 years of UCLO observations of KIC 8462852. Note that the measurements acquired on the C14 West (C14W) have not been renormalised to those on the C14 East (C14E). As in Figure \ref{fig:AllMags}, the errors are the orthogonal sum of the standard error on the nightly averages with the $\sigma_i$ statistic, the derivation of which is described in Section \ref{sec:n2n}. The full table is available for download in machine-readable form at https://figshare.com.}
    \label{AllMagsTab}
    
\end{table*}

\begin{figure*}[!htpb]
	\includegraphics[width=\textwidth]{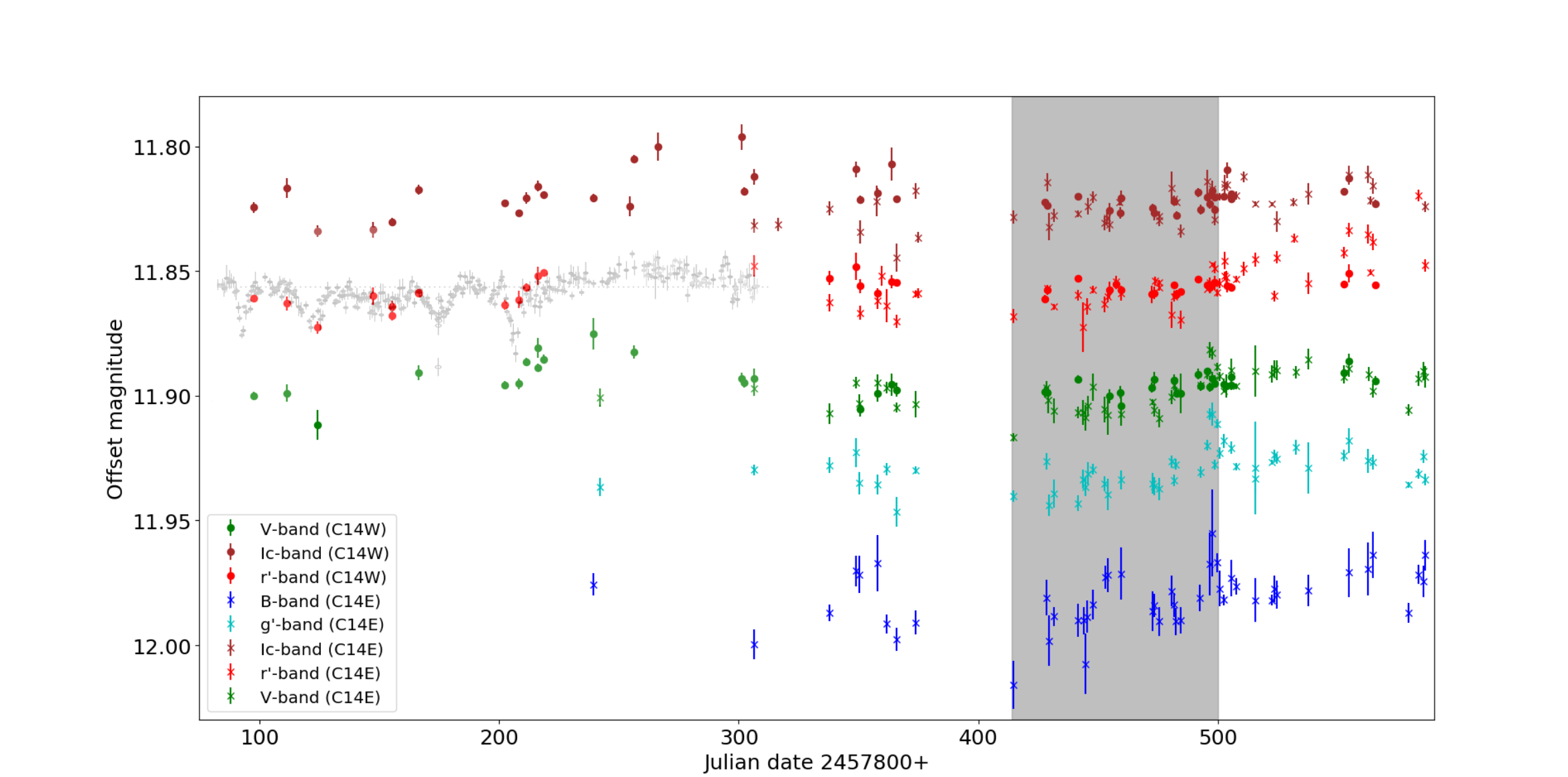}
    \caption{The 1.3-year-long UCLO \textit{$BVg'r'I_c$} light curves for the target star, KIC 8462852. The magnitude time series is characterized by wavelength-dependent dimming and brightening of variable timescales. The gray shaded region describes the Julian date (JD) window (2457800+) 415-500, i.e., the interval analyzed in this study. The May--September 2017 dip events cover the  80--220 interval, over which we superimpose the $r'$-band light curve from Figure 2 of \citet{boyajian2018first}. The \textit{Caral-Supe} and \textit{Evangeline} dips occur in the interval 390-430. Note that for purposes of display, the renormalised $B$, $g'$, $r'$ and $I_c$ passband magnitudes have been adjusted by $-0.43$, $-0.16$, $+0.11$ and $+0.61$ magnitudes respectively. The error bars shown are the orthogonal sum of the standard error on the nightly averages and the $\sigma_{i}$ statistic described in \ref{sec:n2n}. We denote data taken on the C14 West (C14W) and East (C14E) with filled circles and crosses respectively.}
    \label{fig:AllMags}
\end{figure*}

\subsection{Measurements of extinction in the Julian date window 2458215-300}
\label{sec: extinction 8200-8300}

\begin{figure}[!htpb]
\centering
\subfigure{
\label{fig:first}
\includegraphics[height=1.9in]{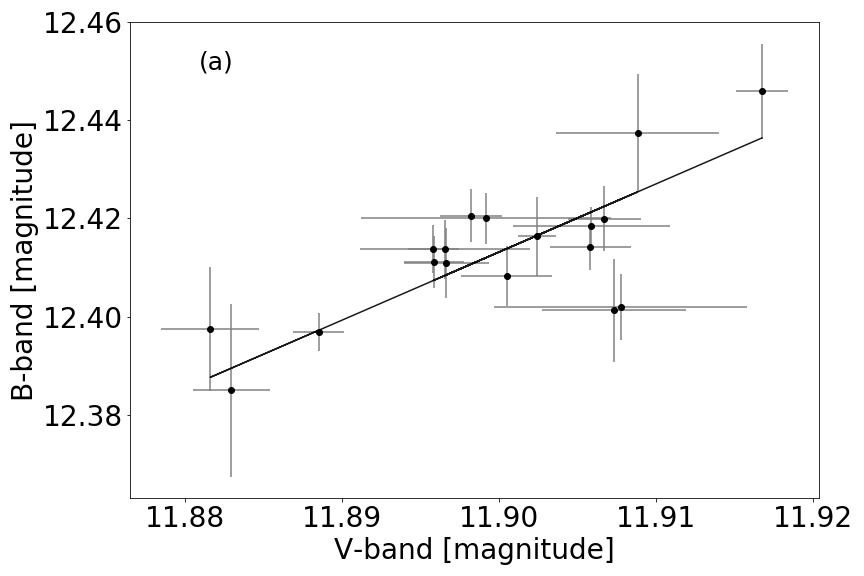}}
\qquad
\subfigure{
\label{fig:second}
\includegraphics[height=1.9in]{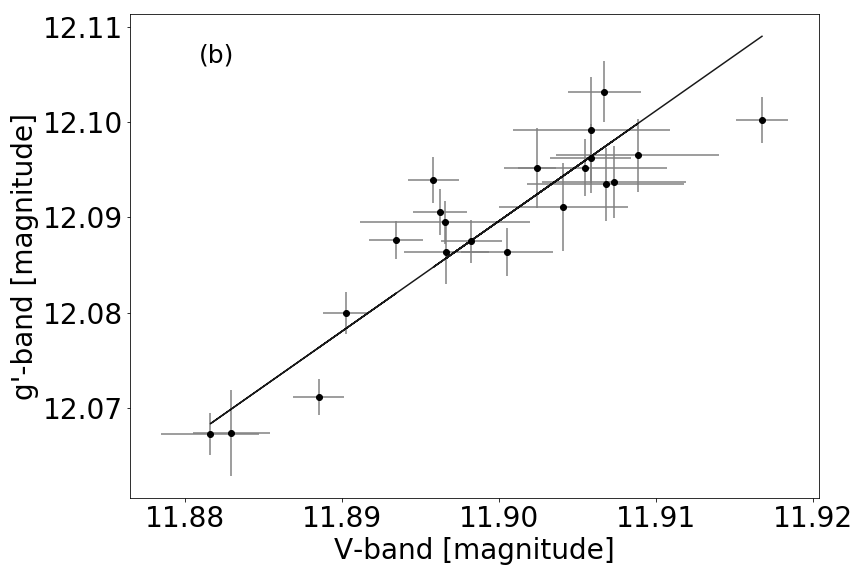}}
\qquad
\subfigure{
\label{fig:third}
\includegraphics[height=1.9in]{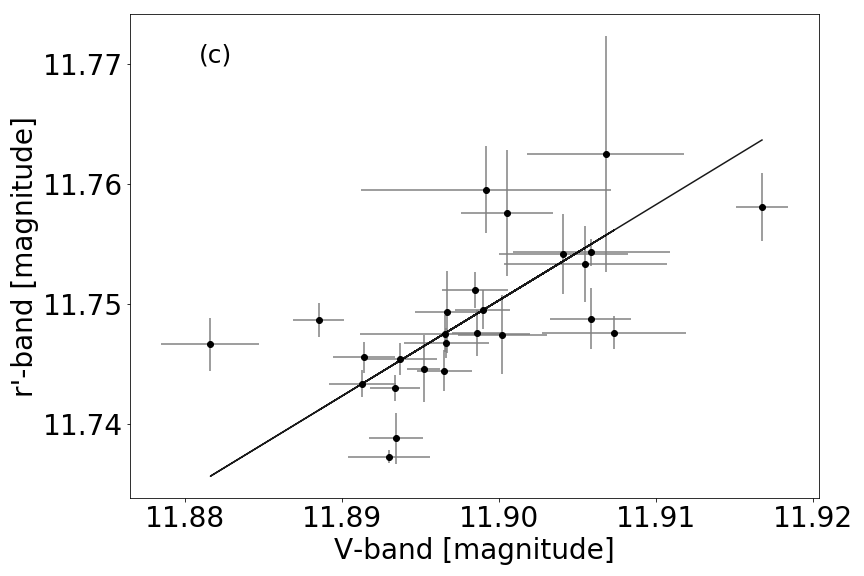}}
\qquad
\subfigure{
\label{fig:fourth}
\includegraphics[height=1.9in]{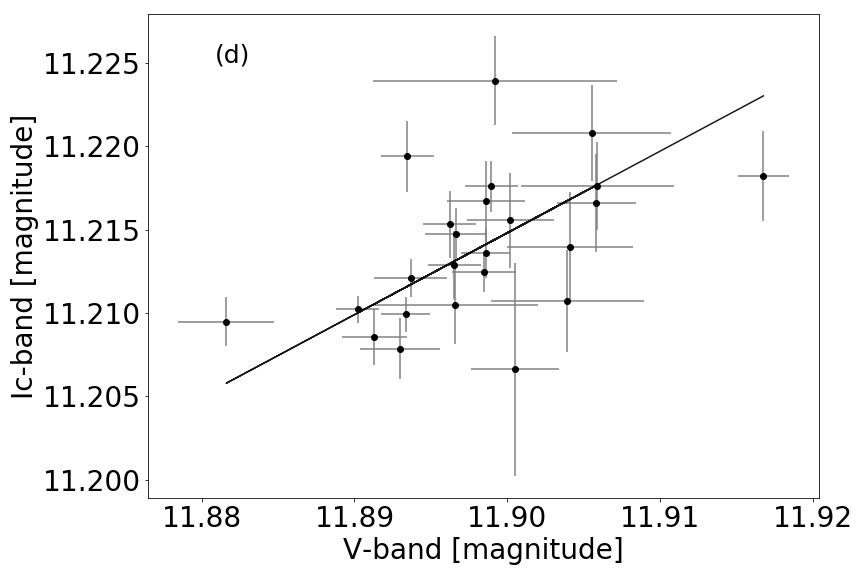}}
\caption{The nightly average magnitudes in $(a)$ $B$, $(b)$ $g'$, $(c)$ $r'$, and $(d)$ $I_C$ passbands against the $V$-band magnitude of KIC 8462852, for the window JD 2458215--300; the fitted lines from a BCES orthogonal weighted least-squares fit are shown.}
\label{fig:RegPlots}
\end{figure}

For each mean magnitude in the combined C14 East and West $V$-band data sets, the closest temporal measurement in the other passbands is found. Provided these measurements were acquired within 0.1 days of each other, the data are considered to be concurrent and matched. On nights where both the C14 West and C14 East were observing, a single nightly average magnitude in passband $X$ (where $X = B$, $g'$, $r'$, or $I_C$) may be matched against both the $V$-band measurements obtained by both telescopes. To avoid `doubling-up' of data, only the closest temporal match is kept.

A measure of the relative extinction  between the different passbands, assumed responsible for these variations, can be determined from a linear fit to the magnitudes obtained in one passband relative to another. Fig. \ref{fig:RegPlots} shows weighted linear fits to the nightly average magnitudes in $B,g',r'$ and $I_C$ passbands against the $V$-band magnitude for the JD window 2458215--300.  

In order to account for the intrinsic scatter and the heteroscedastic measurement errors on both axes in the regression plots (i.e., the size of the errors is dependent on systematic effects that vary from one observation to the next), a BCES (bivariate correlated errors and intrinsic scatter) fitting routine was employed (\citet{akritas1996linear}); an orthogonal least-squares fit is used. The measured slopes, which are equivalent to the ratios of the relative extinctions, $A_\lambda / A_{{V}}$, and shown in Fig. \ref{fig:RegPlots}, are given in Table~\ref{table:extinctions}; we adopt the convention of a normalised measurement of extinction relative to the $V$-band, $A_\lambda/A_V$ (where $\lambda = B$, $g'$, $r'$, or $I_C$). The relative extinctions were then used to fit the function,
\begin{equation}
	\tau_\lambda / \tau_{\lambda_{V}} = (\lambda / \lambda_V)^{-\alpha},
    \label{eq:ang}
\end{equation}
where the optical-depth ratios, $\tau_\lambda / \tau_{\lambda_V}$ can be assumed to be equal to the slopes $A_\lambda / A_V$ if $A_\lambda << 1$, $\lambda$ is the effective passband wavelength, and $\alpha$ is the Angstrom
absorption coefficient
(\citet{moosmuller2011absorption}, \citet{deeg2018non}). From a weighted least-squares fit to the data we infer $\alpha=1.33\pm0.43$. This best fit result is corroborated by a Monte Carlo resampling and refitting of the Table~\ref{table:extinctions} values, with comparable, and normally distributed (i.e., symmetric) uncertainties.

\begin{table}[!htpb]
    \centering
    \begin{tabular}{c  c}
     \hline\hline
    Extinction ratio  & Slope and uncertainty \\ [0.5ex] 
     \hline
    $A_B/A_V$ & $1.39\pm0.27$\\
    $A_{g'}/A_V$ & $1.16\pm0.11$\\
    $A_{r'}/A_V$ & $0.80\pm0.25$\\
    $A_{Ic}/A_V$ & $0.49\pm0.19$\\ [1ex]
    \hline
    \end{tabular}
    \caption{The slopes from orthogonal least-squares regression of nightly average magnitudes in $B,g',r',I_C$ passbands against the $V$-band magnitude of KIC 8462852, for the window JD 2458215--300.}
    \label{table:extinctions}
\end{table}

An approach to minimise the $\chi^2$ of the fit to equation ~\ref{eq:ang}, which simultaneously confers a significant net improvement to the reduced chi-squared of the linear fits in Fig.~\ref{fig:RegPlots}, informed the rejection of a single nightly average magnitude.

\subsection{Comparison with measurements of extinction associated with the month- to year-long brightness variation}

We firmly detect a non-grey extinction for the $\sim$month-long variation seen across the window studied here, and the inferred $\alpha\sim1.3$ --- which translates to a preferred value of $R_V \sim 3.2$ (see below) --- is suggestive of the steeper extinction also measured by \cite{schaefer2018kic}. This is consistent with extinction associated with dust composed of a substantial fraction of sub-micron-sized particles.

The first measurements of the extinction  associated with KIC 8462852's long-term flux variations favoured a more neutral extinction (\S\ref{sec: introduction}), with $R_V \sim 5$, implying larger grain sizes. Using an approach of $\chi^2$ minimisation, \citet{deeg2018non} translate the $R_V$ values of \citet{meng2017extinction} and \citet{davenport2018galex} to $\alpha \leqslant 1.1$ and $\alpha=1.1\pm0.1$ respectively}. With our $\alpha=1.33\pm0.43$, we cannot therefore firmly rule-out this more neutral extinction, even at $1\sigma$. One can also reverse the approach of \citet{deeg2018non} to convert any given $\alpha$ to a corresponding $R_V$, and for our $\alpha=1.33 \pm 0.43$, we recover $R_V = 3.2^{+2.6}_{-1.0}$. As a check, from a weighted least-squares fitting of the values in Table~\ref{table:extinctions} directly to the one-parameter extinction law of \citet{cardelli1989relationship}, with a Monte Carlo resampling based on the uncertainties given, we infer a median $R_V=2.9$ with lower and upper 34 percentiles of $R_V = 1.9$ and $5.5$: i.e., $R_V = 2.9^{+2.6}_{-1.0}$, similar to the values obtained from our fit to $\alpha$.

Despite the large range in $R_V$, it is interesting that our favoured value is biased towards steeper extinctions, similar to that which is associated with material in the interstellar medium. Perhaps crucially, the studies of \citet{meng2017extinction} and \citet{davenport2018galex} relied on measurements taken hundreds of days apart, while this study, as in \citet{schaefer2018kic}, has used data with a time resolution of tens of days or finer. \citet{meng2017extinction} suffer from $\sim100$-day gaps in their \textit{Swift, Spitzer} and \textit{BVR} observations. The \textit{BVRI} light curves of \citet{schaefer2018kic} reveal, however, that there is not a monotonic decline in the interval of these observations --- that is, the observations have bypassed a region showing a steep rise and subsequent decline in the $B$ and $V$ bands of over 10 mmag in the space of $\sim150$ days.

\subsection{The relationship between the dips and the longer-term variation}

With the notable exception of $\alpha=2.19 \pm 0.45$ from \citet{deeg2018non}\footnote{\citet{deeg2018non} do not convert their $\alpha$ to a corresponding $R_V$, but by using the approach described above it is implied to be $R_V \leqslant 2.2$, and hence, representative of a highly chromatic extinction. We note that our upper 1-$\sigma$ limit for $\alpha$ is just consistent with their lower 1-$\sigma$ limit.}, measurements of the extinction associated with dips tend towards more neutral colours than that favoured both here and by \citet{schaefer2018kic} for the longer-term variation. E.g., $R_V \sim 5$ from relative depths $B / i' = 1.94 \pm 0.06$ from \citet{boyajian2018first} (see Section 6 in \cite{schaefer2018kic}), and comparable $(X/i')_{dip}$ depth ratios in \citet{bodman2018variable} (see Table 1 in their Section 3). The exception to this trend is \textit{Celeste}, where \citet{bodman2018variable} measure $B/i' = 3.09^{+0.18}_{-0.17}$ and $r'/i' = 1.55 \pm 0.10$, which together are suggestive of steeper extinctions, such as those favoured in this work.

There is no doubt that the dust responsible for both the day-long dips and month-to-year-long brightness variations of KIC 8462852 must be composed of a large fraction of sub-micron-sized particles. This is highly suggestive of some common origin for these phenomena, which differ only in timescale. In this `single mechanism' framework, the dips may be thought of as the substructure to the longer-term variation \citep{schaefer2018kic}. Of the existing theories attempting to explain this enigmatic star's behaviour, is a contrast in extinction between the dips and longer-term variation to be expected?

The `cometary' hypothesis, first described by \cite{boyajian2016planet}, postulates that the dips can be explained by the passage of a necessarily large number of objects on an highly elliptical orbit that have begun to break-up and release dust (due either to gravitational or thermal stresses) at a periastron close to our line of sight. This theory satisfies both the orbital constraints deduced from the star's light curve, in addition to the apparent absence of an infrared excess typically associated with dusty circumstellar distributions \citep{marengo2015kic,thompson2016constraints}, since the orbital material is at almost all times far from the star, until its approach to its transit. The JD interval assessed in this work immediately follows a large dip event, and hence, in this cometary scenario, would provide a measure of extinction associated with the material composing the `tail' of the fragmenting object(s). \citet{wyatt2017modelling} show how the measured chromaticity is strongly dependent on the optical depth, and indeed, one would associate a steeper extinction with the thinner trailing material than that of the dense dust distribution about the nucleus, which constitutes the cometary coma. One should also expect larger grains to be more closely bound to the nucleus, further influencing the more neutral extinction.

\section{Conclusions}
\label{sec: conclusions}

We present an analysis of the results of a multi-band photometric monitoring campaign of the enigmatic star, KIC 8462852, across the JD window (2450000+) 8215--8300. We measure the extinction associated with the long-term, wavelength-dependent brightness variations of the star seen over this 85-day interval of net brightening, which immediately follows a $\sim5$ per cent days-long dip. We find a significant chromatic dependence in the extinction; we infer an Angstrom absorption coefficient, $\alpha$, of $1.33\pm0.43$ ($R_V \simeq 3.2^{+2.6}_{-1.0}$) to describe the extinction associated with this JD window, which is suggestive of a dust population composed of a substantial fraction of sub-micron-sized grains. This steep extinction is supportive of the most recent analysis of the wavelength-dependence associated with the longer-term brightness variations of the star \citep{schaefer2018kic}, which together are suggestive of a typically stronger extinction than that associated with the days-long dips. That both timescales of brightness variation are associated with sub-micron sized grains suggests they may have a common origin. We cannot, however, confidently rule-out a greyer extinction ($R_V\geqslant5$), even at $1\sigma$, but our fitted values for $\alpha$ and $R_V$ are biased towards a steep extinction, similar to that which is associated with material in the interstellar medium. 

These small sub-micron sized grains would be blown out of the KIC 8462852 system on a timescale of months by the radiation pressure of the F-type star. A consequence of this --- given the reality of the centuries-long \citet{schaefer2016kic} dimming --- is that there must be some continually replenishing source of dust. If these reservoirs of dust are the objects associated with the dips, then further interrogation of the chromatic extinction associated with them may help clarify their nature. It will be important to continue to explore contrasts between the extinction associated with the dips and the longer-term variation through precise multi-color photometry, from which one may begin to assess the relationship between the dust and its progenitors.

\acknowledgments

We thank Mick Pearson and Thomas Schlichter for ensuring continuous and reliable operation of robotic-telescope facilities at UCLO. We would also like to thank Dr. Tabetha Boyajian for allowing us to make use of Figure 2 from \cite{boyajian2018first}, and we extend a thank you to the supporters of her dedicated observing campaign aimed at this most mysterious star.




\bibliography{TabbyStar}{} 

\begin{thebibliography}{}
\expandafter\ifx\csname natexlab\endcsname\relax\def\natexlab#1{#1}\fi
\providecommand{\url}[1]{\href{#1}{#1}}

\bibitem[{Akritas \& Bershady(1996)}]{akritas1996linear}
Akritas, M.~G., \& Bershady, M.~A. 1996, arXiv preprint astro-ph/9605002

\bibitem[{Bertin \& Arnouts(1996)}]{bertin1996sextractor}
Bertin, E., \& Arnouts, S. 1996, Astronomy and Astrophysics Supplement Series,
  117, 393

\bibitem[{Bodman {et~al.}(2018)Bodman, Wright, Boyajian, \&
  Ellis}]{bodman2018variable}
Bodman, E., Wright, J., Boyajian, T., \& Ellis, T. 2018, arXiv preprint
  arXiv:1806.08842

\bibitem[{Boyajian {et~al.}(2016)Boyajian, LaCourse, Rappaport, Fabrycky,
  Fischer, Gandolfi, Kennedy, Korhonen, Liu, Moor,
  {et~al.}}]{boyajian2016planet}
Boyajian, T., LaCourse, D., Rappaport, S., {et~al.} 2016, Monthly Notices of
  the Royal Astronomical Society, 457, 3988

\bibitem[{Boyajian {et~al.}(2018)Boyajian, Alonso, Ammerman, Armstrong, Ramos,
  Barkaoui, Beatty, Benkhaldoun, Benni, Bentley, {et~al.}}]{boyajian2018first}
Boyajian, T.~S., Alonso, R., Ammerman, A., {et~al.} 2018, The Astrophysical
  Journal Letters, 853, L8

\bibitem[{Cardelli {et~al.}(1989)Cardelli, Clayton, \&
  Mathis}]{cardelli1989relationship}
Cardelli, J.~A., Clayton, G.~C., \& Mathis, J.~S. 1989, The Astrophysical
  Journal, 345, 245

\bibitem[{Davenport {et~al.}(2018)Davenport, Covey, Clarke, Laycock, Fleming,
  Boyajian, Montet, Shiao, Million, Wilson, {et~al.}}]{davenport2018galex}
Davenport, J.~R., Covey, K.~R., Clarke, R.~W., {et~al.} 2018, The Astrophysical
  Journal, 853, 130

\bibitem[{Deeg {et~al.}(2018)Deeg, Alonso, Nespral, \& Boyajian}]{deeg2018non}
Deeg, H.~J., Alonso, R., Nespral, D., \& Boyajian, T.~S. 2018, Astronomy \&
  Astrophysics, 610, L12

\bibitem[{Henden {et~al.}(2016)Henden, Templeton, Terrell, Smith, Levine, \&
  Welch}]{henden2016vizier}
Henden, A., Templeton, M., Terrell, D., {et~al.} 2016, VizieR Online Data
  Catalog, 2336

\bibitem[{Lahey {et~al.}(2017)Lahey, Dimick, \& Layden}]{lahey2017improving}
Lahey, A., Dimick, D., \& Layden, A. 2017, Journal of the American Association
  of Variable Star Observers (JAAVSO), 45, 202

\bibitem[{Marengo {et~al.}(2015)Marengo, Hulsebus, \& Willis}]{marengo2015kic}
Marengo, M., Hulsebus, A., \& Willis, S. 2015, The Astrophysical Journal
  Letters, 814, L15

\bibitem[{Meng {et~al.}(2017)Meng, Rieke, Dubois, Kennedy, Marengo, Siegel, Su,
  Trueba, Wyatt, Boyajian, {et~al.}}]{meng2017extinction}
Meng, H.~Y., Rieke, G., Dubois, F., {et~al.} 2017, The Astrophysical Journal,
  847, 131

\bibitem[{Moosm{\"u}ller {et~al.}(2011)Moosm{\"u}ller, Chakrabarty, Ehlers, \&
  Arnott}]{moosmuller2011absorption}
Moosm{\"u}ller, H., Chakrabarty, R., Ehlers, K., \& Arnott, W. 2011,
  Atmospheric Chemistry and Physics, 11, 1217

\bibitem[{Osborn {et~al.}(2019)Osborn, Kenworthy, Rodriguez, de~Mooij, Kennedy,
  Relles, Gomez, Hippke, Banfi, Barbieri, {et~al.}}]{osborn2019pds}
Osborn, H.~P., Kenworthy, M., Rodriguez, J.~E., {et~al.} 2019, Monthly Notices
  of the Royal Astronomical Society

\bibitem[{Schaefer(2016)}]{schaefer2016kic}
Schaefer, B.~E. 2016, The Astrophysical Journal Letters, 822, L34

\bibitem[{Schaefer {et~al.}(2018)Schaefer, Bentley, Boyajian, Coker, Dvorak,
  Dubois, Erdelyi, Ellis, Graham, Harris, {et~al.}}]{schaefer2018kic}
Schaefer, B.~E., Bentley, R.~O., Boyajian, T.~S., {et~al.} 2018, Monthly
  Notices of the Royal Astronomical Society, 481, 2235

\bibitem[{Thompson {et~al.}(2016)Thompson, Scicluna, Kemper, Geach, Dunham,
  Morata, Ertel, Ho, Dempsey, Coulson, {et~al.}}]{thompson2016constraints}
Thompson, M., Scicluna, P., Kemper, F., {et~al.} 2016, Monthly Notices of the
  Royal Astronomical Society: Letters, 458, L39

\bibitem[{Wyatt {et~al.}(2017)Wyatt, van Lieshout, Kennedy, \&
  Boyajian}]{wyatt2017modelling}
Wyatt, M.~C., van Lieshout, R., Kennedy, G.~M., \& Boyajian, T. 2017, Monthly
  Notices of the Royal Astronomical Society, 473, 5286

\end{thebibliography}








\label{lastpage}

\end{document}